\documentclass[aps,prd,preprint,superscriptaddress]{revtex4-1}
\usepackage[english]{babel}
\usepackage[utf8]{inputenc}
\usepackage[T1]{fontenc}
\usepackage{amsmath,amssymb,braket,slashed,mathrsfs}
\usepackage{pifont}
\usepackage{hyperref}
\usepackage{graphicx}
\graphicspath{{./figures/}}
\usepackage{tikz}
\usetikzlibrary{shapes,arrows,positioning}
\tikzset{decision/.style={diamond, draw, fill=blue!20, text width=4.5em, text badly centered, inner sep=0pt}}
\tikzset{block/.style={rectangle, draw, fill=blue!20, text width=10em, text centered, rounded corners, minimum width=3.5cm}}
\tikzset{block1/.style={rectangle, draw, fill=blue!20, text width=18.5em, text centered, rounded corners, minimum width=3.5cm}}
\tikzset{line/.style={draw, -latex, thick}}
\usepackage{color,hyperref}
\usepackage{cleveref}
\newcommand{\ba}{\begin{eqnarray}}
\newcommand{\ea}{\end{eqnarray}}
\newcommand{\bi}{\begin{itemize}}
\newcommand{\ei}{\end{itemize}}

\newcommand{\MS}{\overline{\mathrm{MS}}}
\newcommand{\lat}{\mathrm{lat}}

\usepackage{multirow,array}

\definecolor{linkcolor}{rgb}{.17578125,.1875,.5703125}
\hypersetup{
    pdfstartview={FitH},    
    pdftitle={Rare kaon decays: a lattice study},    
    pdfauthor={RBC and UKQCD collaborations},           
    pdfsubject={Physics article},                
    pdfcreator={Xu Feng},          
    pdfkeywords={particle} {physics} {quantum} {chromodynamics} {QCD}  {hadrons} {kaons} {rare decays} {flavor} {FCNC} {lattice}, 
    colorlinks=true,        
    linkcolor=linkcolor,    
    citecolor=linkcolor,    
    filecolor=black,        
    urlcolor=linkcolor
}

\newcommand{\cu}{Physics Department, Columbia University, New York, NY 10027, USA}
\newcommand{\soton}{Department of Physics and Astronomy, University of
Southampton, Southampton SO17 1BJ, UK}
\newcommand{\pkuphy}{School of Physics and State Key Laboratory of Nuclear
Physics and Technology, Peking University, Beijing 100871, China}
\newcommand{\innovation}{Collaborative Innovation Center of Quantum Matter, Beijing 100871, China}
\newcommand{\chep}{Center for High Energy Physics, Peking University, Beijing 100871, China}
\newcommand{\uoe}{School of Physics and Astronomy, University of Edinburgh, Edinburgh EH9 3JZ, UK}

\begin{document}
\title{Lattice QCD study of the rare kaon decay
\boldmath{$K^+\to\pi^+\nu\bar{\nu}$} at a near-physical pion mass}

\author{Norman~H.~Christ}\affiliation{\cu}
\author{Xu~Feng}\affiliation{\pkuphy,\innovation,\chep}
\author{Antonin~Portelli}\affiliation{\uoe}
\author{Christopher~T.~Sachrajda}\affiliation{\soton}
\collaboration{RBC and UKQCD collaborations}
\pacs{PACS}

\date{\today}

\begin{abstract}
    The rare kaon decay $K^+\to\pi^+\nu\bar{\nu}$ is an ideal process in which 
to search for signs of new physics and is the primary goal of the NA62 experiment at
    CERN. In this paper we report on a lattice QCD calculation of the 
    long-distance contribution to the $K^+\to\pi^+\nu\bar{\nu}$ decay amplitude at the near-physical 
    pion mass $m_\pi=170$ MeV. The calculations are however, performed on 
    a coarse lattice and hence with a lighter charm quark mass ($m_c^{\MS}(\mbox{3\,GeV})=750$\,MeV) than the physical one. The main aims of this study are two-fold. 
    Firstly we study the momentum dependence of the amplitude and conclude that it is very mild so that a computation at physical masses even at a single kinematic point would provide a good estimate of the long-distance contribution to the decay rate. 
Secondly we compute the contribution to the branching ratio from the two-pion intermediate state whose energy is below the kaon mass and find that it is less than 1\%\, after its exponentially growing unphysical 
contribution has been removed and that the 
corresponding non-exponential finite-volume effects are negligibly small.
\end{abstract}

\maketitle


\section{Introduction}\label{sec:intro}

The rare kaon decays $K\to\pi\nu\bar{\nu}$ have attracted increasing interest during the past few decades. As flavor changing neutral current (FCNC) processes, these decays are highly suppressed in the standard model (SM) and thus provide ideal probes for the observation of new physics. Moreover, these decays are short-distance (SD) dominated and are therefore theoretically clean. In this paper we denote contributions from distances greater than $O(1/m_c)$, where $m_c$ is the mass of the charm quark, as long-distance (LD) contributions. Such LD contributions can safely be neglected in the CP-violating
$K_L\to\pi^0\nu\bar{\nu}$ decays and are expected to be small, perhaps of $O(5$\,-\,$10\%)$,  in $K^+\to\pi^+\nu\bar{\nu}$ decays. 
The study reported in this paper is the next step in our project to compute the LD contributions non-perturbatively using lattice QCD and to obtain the branching ratio for $K^+\to\pi^+\nu\bar{\nu}$ decays with a QCD precision of $O(1\%)$ (uncertainties in the Cabibbo-Kobayashi-Maskawa (CKM) matrix elements need to be reduced in parallel with the reduction of QCD errors). The usual approach simply uses perturbation theory to integrate out the charm quark. An estimate of the missing LD effects in this approach is based on chiral perturbation theory and suggests a $(6\pm 3)$\% enhancement of the branching ratio $\mathrm{Br}(K^+\to\pi^+\nu\bar{\nu})$\,\cite{Isidori:2005xm}, which is comparable to
the 8\% total SM error\,\cite{Buras:2015qea}. We will test this result using lattice QCD.

The current experimental value of the branching ratio, $\mathrm{Br}(K^+\to\pi^+\nu\bar{\nu})=(1.73^{+1.15}_{-1.05})\times 10^{-10}$~\cite{Artamonov:2008qb}, is a combined result based on the 7 events collected by the E787 experiment at the Brookhaven National Laboratory\,\cite{Adler:1997am,Adler:2000by,Adler:2001xv,Adler:2002hy} and its successor experiment E949\,\cite{Anisimovsky:2004hr,Artamonov:2008qb}. This result can be compared to SM theoretical predictions such as that in Ref.\,\cite{Buras:2015qea}, $\mathrm{Br}(K^+\to\pi^+\nu\bar{\nu})_{\mathrm{SM}}=(9.11\pm0.72)\times 10^{-11}$, highlighting the need to reduce both the experimental and theoretical errors. 
The current experiment, NA62 at CERN,
which aims at an observation of $O(100)$ events and a 10\%-precision measurement
of $\mathrm{Br}(K^+\to\pi^+\nu\bar{\nu})$, has set an upper limit of
$14\times10^{-10}$ on the $K^+\to\pi^+\nu\bar{\nu}$ branching ratio at 95\%
C.L.~\cite{CortinaGil:2018fkc}.
Since it will be possible to compare the SM predictions with the new experimental measurement of $\mathrm{Br}(K^+\to\pi^+\nu\bar{\nu})$ relatively soon, 
a lattice QCD calculation of the LD contributions is important and timely.

The KOTO experiment at J-PARC, is designed to search for $K_L\to\pi^0\nu\bar{\nu}$ decays\,\cite{Yamanaka:2012yma}.
It has observed one candidate event based on the first 100 hours of physics running in 2013 and set an upper limit of
 $5.1\times10^{-8}$ for the branching ratio at 90\% confidence level\,\cite{Ahn:2016kja}. For a recent description of the status of this experiment see Ref.\,\cite{Shinohara:2019xxx}.
Since the LD contributions to $K_L\to\pi^0\nu\bar{\nu}$ decays are negligible, we do not discuss these decays further in this paper. 

This paper is the latest in a series of lattice QCD studies of the rare kaon
decays $K\to\pi\ell^+\ell^-$ and $K\to\pi\nu\bar{\nu}$, in which we have
developed the theoretical framework and performed exploratory numerical
calculations\,\cite{Sachrajda:2013vqa,Sachrajda:2013fxa,Feng:2015kfa,Christ:2015aha,Christ:2016eae,Christ:2016psm,Christ:2016awg,Christ:2016lro,Christ:2016mmq,Lawson:2017kxc,Bai:2017fkh,Bai:2018hqu}. Here we focus on the LD contributions to the $K^+\to\pi^+\nu\bar{\nu}$ decay amplitude.
Since we have two neutrinos in the final state and the strangeness changes by one unit, this is a doubly weak process and the long-distance contributions arise either from the exchange of two W-bosons ($W$-$W$ diagrams) or the exchange of one $Z$-boson and one $W$-boson ($Z$-exchange diagrams)\,\cite{Christ:2016eae}. When evaluating the LD contribution of both types of diagram the $W$ and $Z$ exchanges are replaced by four-quark operators.  However, the integral over the locations of these four-quark operators is logarithmically divergent when their locations coincide, requiring the introduction of an additional counter term.
In the full SM these divergences are cut off by the mass of the $W$-boson. To obtain the physical amplitude, we must consistently combine the long distance contributions determined from lattice QCD with the short distance contributions which can be reliably determined using perturbation theory\,\cite{Buchalla:1998ba,Misiak:1999yg,Brod:2010hi,Buras:2005gr,Buras:2006gb,Brod:2008ss}. An exploratory calculation of the $K^+\to\pi^+\nu\bar{\nu}$ decay amplitude has been performed in Refs.\,\cite{Bai:2017fkh,Bai:2018hqu} with meson masses $m_\pi\simeq 420$\,MeV, $m_K\simeq 563\,$MeV and a charm-quark mass $m_c^{\MS}(\mbox{2\,GeV})\simeq 860$\,MeV. Because of the small phase space available with these masses, the allowed momenta for the final-state particles are constrained to lie in a narrow region and provide little information on the momentum dependence of the decay amplitude.
For this reason, we computed the amplitude in Refs.\,\cite{Bai:2017fkh,Bai:2018hqu} at a single choice of momenta. The momentum dependence of the decay amplitude was therefore unresolved and is the focus of the study reported here. 

In this paper we calculate the amplitude at the near-physical pion mass $m_\pi\simeq 170$\,MeV and kaon mass $m_K\simeq 493$\,MeV and study the momentum dependence of the $K\to\pi\nu\bar{\nu}$ decay amplitude. In addition since now, in contrast to the earlier studies in\,Refs.\,\cite{Bai:2017fkh,Bai:2018hqu}, the mass of the kaon is above the two-pion threshold, we are able to study the contribution of the $\pi\pi$ intermediate state to the decay amplitude together with the associated finite-volume effects. The conclusions are presented in Sec.\,\ref{sec:concs} but we anticipate them briefly here to say that we find that the momentum dependence of the LD contributions is very mild and that the contribution of the lowest-energy finite-volume two-pion intermediate state to the branching ratio is at the sub-percent level. 

The plan for the remainder of this paper is as follows. In Section\,\ref{sec:details} we present the details of our computation. 
This is followed by the results for the momentum dependence in Sec.\,\ref{sec:results}. 
The contribution of the lowest-energy, finite-volume two-pion intermediate state, computed at the maximum value of $s=(p_K-p_\pi)^2$, is discussed in Sec.\,\ref{sec:pipi}.
Finally in Sec.\,\ref{sec:concs} we present our conclusions.

\section{Details of the Lattice Calculation}\label{sec:details}
   In this study we use the $2+1$ flavor domain wall fermion configurations with lattice size
   $L^3\times T\times L_s=32^3\times64\times32$ (where $L_s$ is the extent
of the lattice in the fifth dimension)
and
   the Iwasaki+DSDR (dislocation suppressing determinant ratio) gauge action at an inverse lattice spacing $a^{-1}=1.37(1)$\,GeV, generated by the RBC-UKQCD collaborations~\cite{Arthur:2012yc}.
   By using M\"obius valence fermions with $L_s=12$ and the M\"obius parameter $b+c=2.667$, one can fix $L_s\times(b+c)=32$ and maintain the residual mass at its unitary value $am_{\mathrm{res}}=0.001845(6)$.
   The smaller extent of the fifth dimension results in a significantly reduced computational cost.
   The pion and kaon masses are found to be $m_\pi=172(1)$\,MeV and $m_K=493(3)$\,MeV.
   The bare mass of the valence charm quark is $am_c=0.38$ which, combined with the mass renormalization factor
   $Z_m^{\MS}(\mbox{3 GeV})=1.44$\,\cite{Arthur:2012yc}, corresponds to the $\MS$
   mass $m_c^{\MS}(\mbox{3\,GeV})=750$\,MeV. 
   We use 100 configurations,
   each separated by 16 molecular dynamics time units. All the results
   presented below are given in lattice units
   unless otherwise specified.

  The decay amplitude for the rare kaon decay $K^+(p_K)\to\pi^+(p_\pi)\nu(p_\nu)\bar{\nu}(p_{\bar{\nu}})$ can be written as the product of 
  a scalar amplitude $F(s,\Delta)$ and a spinor product $\bar{u}(p_\nu){\slashed p}_K(1-\gamma_5)v(p_{\bar{\nu}})$\,\cite{Christ:2016eae}: 
  \ba\label{eq:Fdef}
  A(K^+\to\pi^+\nu\bar{\nu})=i\, F(s,\Delta)\left[\bar{u}(p_\nu){\slashed p}_K(1-\gamma_5)v(p_{\bar{\nu}})\right].
  \ea
  Here we use Euclidean four-momenta with imaginary time components for on-shell particles, e.g. $p_\pi=(iE_\pi,{\bf p}_\pi)$ for the pion. The Lorentz invariant $s\equiv-(p_K-p_\pi)^2$ is the square of the invariant mass of the $\nu\,\bar{\nu}$ pair  and takes  values in the range  $s\in[0,s_{\mathrm{max}}]$ where $s_{\mathrm{max}}=(m_K-m_\pi)^2$.
The other invariant in Eq.\,(\ref{eq:Fdef}) is $\Delta$, which is defined by $\Delta\equiv(p_K-p_\nu)^2-(p_K-p_{\bar{\nu}})^2$. For each value of $s$, $\Delta$ lies in the range $\Delta\in[-\Delta_{\mathrm{max}}(s),+\Delta_{\mathrm{max}}(s)]$, where $\Delta_{\mathrm{max}}(s)=\sqrt{(m_K^2+m_\pi^2-s)^2-4m_K^2m_\pi^2}$.
We denote by $\Delta_{\mathrm{max}}=m_K^2-m_\pi^2$ the absolute maximum of $\Delta$; this can be reached for $s=0$.
One can also write $\Delta$ as $\Delta=\Delta_{\mathrm{max}}(s)\,\cos\theta$, where $\theta$ indicates the angle between $\vec{p}_\pi$ and $\vec{p}_\nu$ in the rest frame of the $\nu\,\bar{\nu}$ pair. For a particular $K^+\to\pi^+\nu\bar\nu$ event the quantity $\Delta$ cannot be determined experimentally and so a suitable phase-space integral must be performed to obtain either the full rate, or the differential distribution with respect to $s$ (the variable $s$ depends on only the momenta of the charged mesons and is therefore measurable).

Note that the square of the spinor product $|\bar{u}(p_\nu){\slashed p}_K(1-\gamma_5)v(p_{\bar{\nu}})|^2=\Delta_{\mathrm{max}}^2(s)-\Delta^2$, so that at the edge of the allowed momentum region, where $\Delta=\Delta_{\mathrm{max}}(s)$, the decay amplitude vanishes. We are interested therefore in computing the amplitude at momenta away from this kinematic edge.
In our calculation, we compute the amplitude at the following four pairs of $(s,\Delta)$:
\ba
\label{eq:mom_mode}
(s,\Delta)=(0,0),\quad \left(0,\frac{\Delta_{\mathrm{max}}}{2}\right),\quad \left(\frac{s_{\mathrm{max}}}2,0\right),\quad \left(\frac{s_{\mathrm{max}}}3,\frac{\Delta_{\mathrm{max}}}3\right).
\ea 
The scalar amplitude $F(s,\Delta)$ is computed for both the $W$-$W$ and $Z$-exchange diagrams. We denote the contribution from the $W$-$W$ diagrams by $F_{WW}(s,\Delta)$; as indicated this depends on both $s$ and $\Delta$. We denote the contribution from the $Z$-exchange diagrams by $F^Z_+(s)$ which, again as indicated, 
depends on only the single variable $s$ since the neutrino and antineutrino are emitted from a single point. The quantity $F^Z_+(s)$ is analogous to the corresponding form factor in $K_{\ell 3}$ decays, but with the charged weak current replaced by the neutral one.
For massless neutrinos, the second form factor $F_-^Z(s)$ does not contribute to the amplitude.  We calculate the $Z$-exchange diagram at one additional kinematic point $(s,\Delta)=(s_{\mathrm{max}},0)$. Although at this point  $F^Z_+(s)$ cannot be determined directly, we can obtain the amplitude for $F^Z_0(s)=F_+^Z(s)+\frac{s}{M_K^2-M_\pi^2}F_-^Z(s)$. If, as expected, $F_-^Z(s)$ is significantly smaller than $F_+^Z(s)$, then $F^Z_0(s)$ is a good approximation to $F_+^Z(s)$.
For a complete explanation on how we compute the scalar amplitudes $F^{WW}(s,\Delta)$, $F_\pm^Z(s)$ and $F_0^Z(s)$, we refer the readers to Refs.\,\cite{Christ:2016eae,Bai:2017fkh,Bai:2018hqu}.

\section{Momentum dependence of the amplitude}\label{sec:results}
For the $Z$-exchange diagrams, we calculate the scalar amplitudes $F_{+,-,0}^{Z}(s)$ at 
$s=0$, $s_{\mathrm{max}}/3$ and $s_{\mathrm{max}}/2$ and present the results in Table\,\ref{tab:lattice_results_Z_exchange}. 
For $s=0$, we have $F_{+}^{Z}(s)=F_{0}^Z(s)$.
For $s=s_{\mathrm{max}}/3$ and $s_{\mathrm{max}}/2$, we find that the values of $F_0^Z(s)$ are a little smaller than those of $F_+^Z(s)$. However, within the statistic
uncertainties, $F_0^Z(s)$ and $F_+^Z(s)$ are consistent, suggesting that $F_{0}^Z(s)$ is a good approximation for $F_{+}^{Z}(s)$.

\begin{figure}[t]
\begin{center}
\includegraphics[width=0.35\hsize]{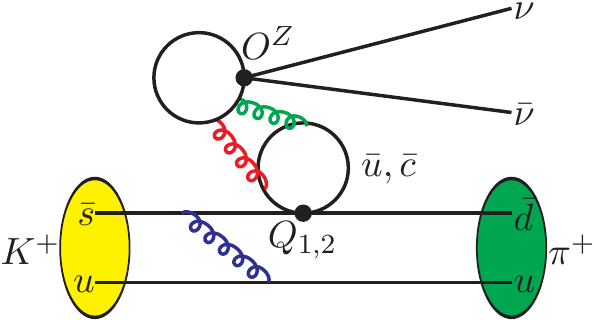}\qquad\qquad
\includegraphics[width=0.35\hsize]{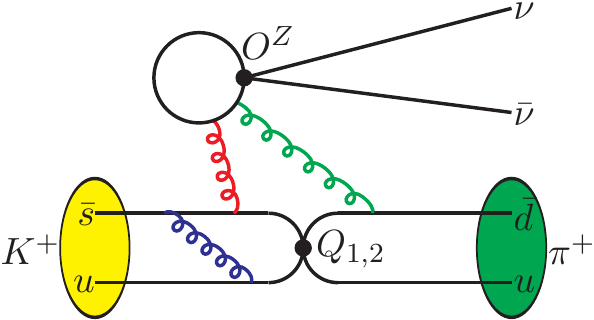}
\end{center}
\caption{Disconnected $Z$-exchange diagrams. Such diagrams cannot be evaluated with the twisted boundary conditions for the valence $\bar{d}$ antiquark which are used for the evaluation of $F_{+,-,0}(s)$ at $s=0,\,s_\mathrm{max}/3$ and $s_\mathrm{max}/2$.}
\label{fig:disconnectedZ}
\end{figure}
\begin{figure}
    \centering
\includegraphics[width=0.8\textwidth]{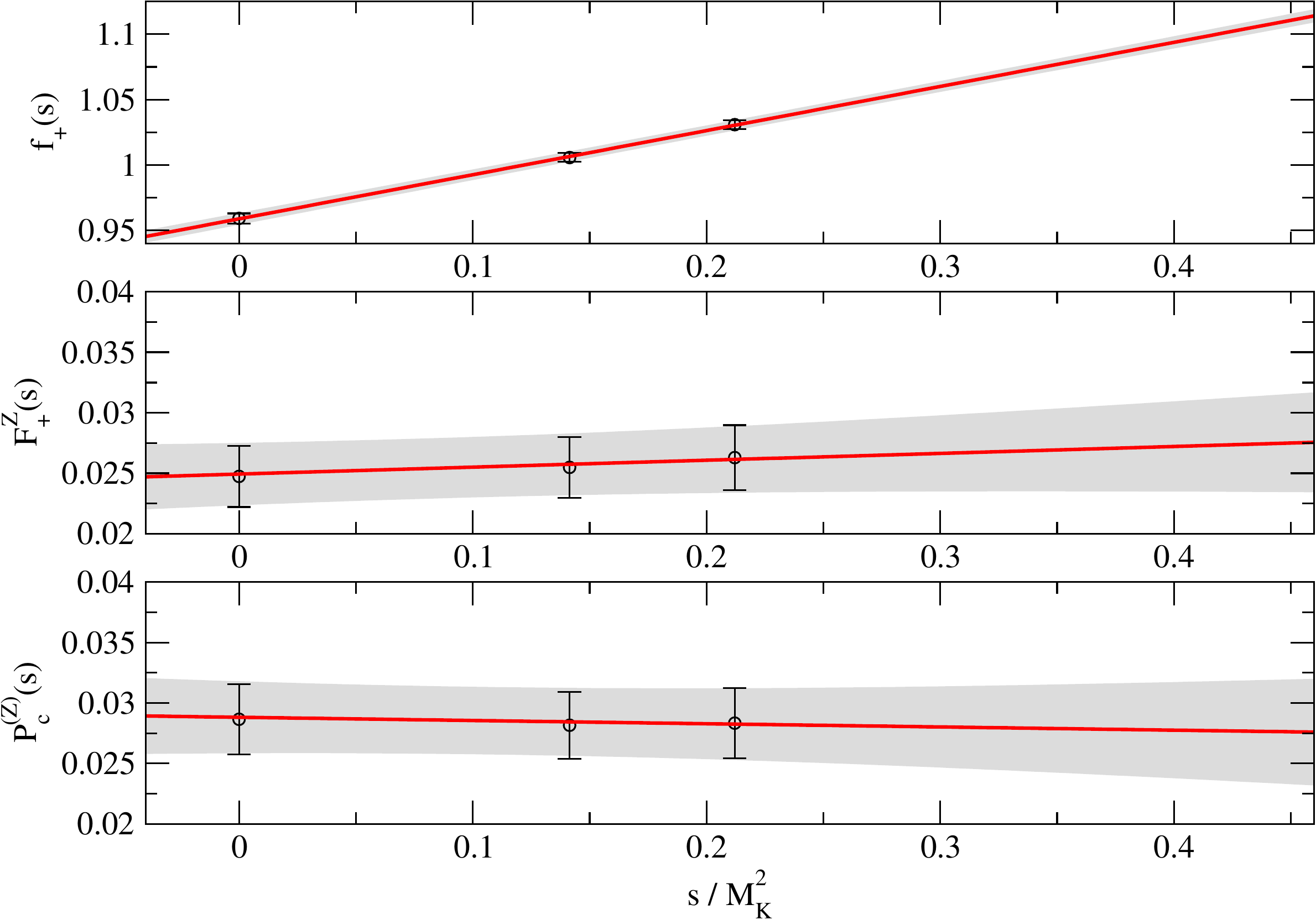}
   \caption{From top to bottom: $f_+(s)$, $F_+^Z(s)$ and $P_c^{(Z)}(s)$ as functions of $s$.}
    \label{fig:mom_dep_Z}
    \end{figure}

In addition we evaluate the scalar amplitude $F_{0}^{Z}(s)$ at $s=s_{\mathrm{max}}$. At this kinematic point the pion and kaon are both at rest and so we can only determine a single form factor, $F_{0}^Z(s_\mathrm{max})$.
The main motivation for the computation of $F_{0}^{Z}(s_\mathrm{max})$ is to provide an estimate of the contribution of the 
disconnected diagrams. At the other kinematic points we use twisted boundary conditions to obtain results for a range of momenta, all with $s\ge 0$, whereas at $s=s_\mathrm{max}$ we use periodic boundary conditions. With twisted boundary 
conditions for the valence down quark, the neutrinos carry momenta which are not integer multiples  of $2\pi/L$, and such momenta cannot be transferred by the gluons which satisfy periodic boundary conditions (see Fig.\,\ref{fig:disconnectedZ}). 
From the results in Table\,\ref{tab:lattice_results_Z_exchange} we conclude that 
the disconnected diagrams make only a small contribution to
$F_{0}^{Z}(s_\mathrm{max})$ and since we expect $F_+^Z(s)\simeq F_0^Z(s)$ and
have found the momentum dependence of $F_+(s)$ to be very mild (see
Fig.\,\ref{fig:mom_dep_Z} and the discussion below), this suggests that the
contribution of the disconnected diagrams to the $K^+\to\pi^+\nu\bar{\nu}$ decay
amplitude is much less than 1\%.

\begin{table}
\begin{tabular}{l c | c | c}
$Z$-exchange & $Q_1$ & $Q_2$ & $C_1^{\lat}Q_1+C_2^{\lat}Q_2$ \\
\hline\hline
\multicolumn{3}{l}{$K^+(\vec 0\hspace{1pt})\to\pi^+(\vec{p}\hspace{1pt})$, $s=0$} \\
\hline

$F_+^{Z}(0)=F_0^{Z}(0)$ 
& $-11.86(37)\cdot10^{-2}$ & $-0.19(32)\cdot10^{-2}$ & $2.47(25)\cdot10^{-2}$ \\
$F_-^{Z}(0)$ 
& $1.81(83)\cdot10^{-2}$ & $-1.04(89)\cdot10^{-2}$ & $-1.02(66)\cdot10^{-2}$ \\
\hline
\hline
\multicolumn{3}{l}{$K^+(\vec 0\hspace{1pt})\to\pi^+(\vec p\hspace{1pt})$, $s=s_{\mathrm{max}}/3$} \\
\hline

$F_+^{Z}(s)$ 
& $-12.56(39)\cdot10^{-2}$ & $-0.31(32)\cdot10^{-2}$ & $2.55(25)\cdot10^{-2}$ \\
$F_-^{Z}(s)$ 
& $2.13(81)\cdot10^{-2}$ & $-0.68(83)\cdot10^{-2}$ & $-0.87(63)\cdot10^{-2}$ \\
$F_0^{Z}(s)$
& $-12.22(30)\cdot10^{-2}$ & $-0.42(24)\cdot10^{-2}$ & $2.41(19)\cdot10^{-2}$ \\
\hline
\hline
\multicolumn{3}{l}{$K^+(\vec 0\hspace{1pt})\to\pi^+(\vec p\hspace{1pt})$, $s=s_{\mathrm{max}}/2$} \\
\hline

$F_+^{Z}(s)$ 
& $-13.10(40)\cdot10^{-2}$ & $-0.38(34)\cdot10^{-2}$ & $2.63(27)\cdot10^{-2}$ \\
$F_-^{Z}(s)$ 
& $2.55(82)\cdot10^{-2}$ & $-0.46(86)\cdot10^{-2}$ & $-0.83(65)\cdot10^{-2}$ \\
$F_0^{Z}(s)$
& $-12.49(27)\cdot10^{-2}$ & $-0.49(21)\cdot10^{-2}$ & $2.43(16)\cdot10^{-2}$ \\
\hline\hline
\multicolumn{3}{l}{$K^+(\vec 0\hspace{1pt})\to\pi^+(\vec 0\hspace{1pt})$, $s=s_{\mathrm{max}}$}\\ 
\hline

$F_0^{Z}(s)$ 
& $-12.75(19)\cdot10^{-2}$ & $-1.20(16)\cdot10^{-2}$ & $2.05(12)\cdot10^{-2}$ \\

$F_0^{Z,\mathrm{disc}}(s)$ 
& $-19.1(6.8)\cdot10^{-4}$ & $0.9(6.2)\cdot10^{-4}$ & $4.7(4.8)\cdot10^{-4}$  \\
\hline
\end{tabular}
\caption{Results for the form factors $F^Z_{+/-/0}$ from the
$Z$-exchange diagrams. From top to bottom, we show the results for the momentum transfer $s=0$, $s_{\mathrm{max}}/3$, $s_{\mathrm{max}}/2$ and $s=s_{\mathrm{max}}$, respectively,
where $s_{\mathrm{max}}=(m_K-m_\pi)^2$. $Q_{1,2}$ are the conventional $\Delta S=1$ bare lattice operators defined in Eq.\,(15) of Ref.\,\cite{Christ:2016eae} and the columns labeled by $Q_{1,2}$ contain the results with $Q_{1}$ or $Q_2$ as the $\Delta S=1$ operator. In the final column we combine these results with the corresponding Wilson coefficients $C_{1,2}$. All results are in lattice units.}
\label{tab:lattice_results_Z_exchange}
\end{table}

The momentum dependence of the form factors is plotted in Fig.\,\ref{fig:mom_dep_Z}. In the upper panel, we show the form factor $f_+(s)$ for the $K^+\to\pi^0\ell^+\nu$ ($K_{\ell 3}$) decay. In the middle panel, we present $F_+^Z(s)$ as a function of $s/m_K^2$. Finally, in the lower panel $F_+^Z(s)$ is normalized to convert it 
into the conventional phenomenological quantity $P_c^{(Z)}$ defined as (see Eq.\,(88) in Ref.\,\cite{Bai:2018hqu}):
\ba
P_c^{(Z)}(s)\equiv \frac{\pi^2}{\lambda^4M_W^2}\frac{F_+^Z(s)}{f_+(s)}\,.
\ea
Here we use a superscript {\footnotesize $(Z)$} to indicate that this is the contribution from the $Z$-exchange diagrams. In order to obtain the full contribution of the $Z$-exchange diagrams to $F_+^Z(s)$, we need to subtract the short-distance counter-term from $C_1^\mathrm{lat}Q_1+C_2^\mathrm{lat}Q_2$, the combination given in the last column of Table\,\ref{tab:lattice_results_Z_exchange}, following the procedure described in detail in Refs.\,\cite{Christ:2016eae,Bai:2018hqu}.
This short-distance contribution to $F_+^Z(s)$, which is difficult to compute on such a coarse lattice, is proportional to $f_+(s)$ and hence appears as a constant term in $P_c^{(Z)}(s)$. 
Since the focus of this paper is to explore the $s$-dependence of $P_c^{(Z)}(s)$, we do not attempt to perform this subtraction.
From the bottom panel in Fig.\,\ref{fig:mom_dep_Z}, we conclude that the momentum dependence of $P_c^{(Z)}$ is very mild, and cannot be resolved within the precision of our calculation. To quantify this statement we parametrize the $s$-dependence of $P_c^{(Z)}(s)$ by the linear function 
\ba
P_c^{(Z)}(s)=P_c^{(Z)}(0)+b_s^{(Z)}\frac{s}{m_K^2}
\ea
and find $b_s^{(Z)}=-(1.8\pm 9.7)\times10^{-3}$ from a correlated fit.

   \begin{figure}
    \centering
         \shortstack{
     \shortstack{\includegraphics[width=.3\textwidth]{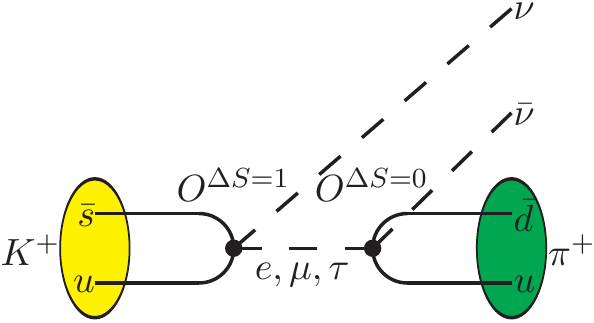}\\Type 1}
     \hspace{0.3cm}
         \shortstack{\includegraphics[width=.3\textwidth]{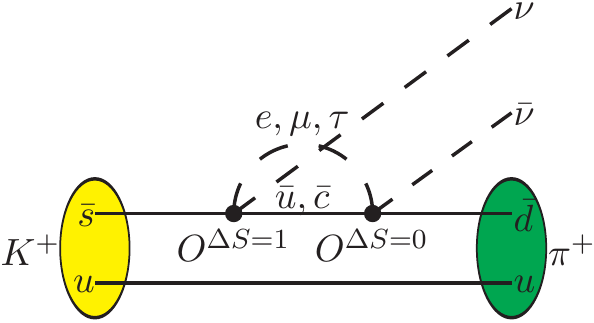}\\Type 2}
     \\$W$-$W$ diagram}
   \caption{Quark and lepton contractions for the $W$-$W$ diagrams.}
    \label{fig:contraction}
    \end{figure}

For the $W$-$W$ diagrams, we calculate the scalar amplitude $F_{WW}$ for the momentum pairs $(s,\Delta)$ given in Eq.\,(\ref{eq:mom_mode}). This scalar amplitude is divided into two parts corresponding to the different contractions labelled as Type\,1 and Type\,2 in Fig.\,\ref{fig:contraction}. The $W$-$W$ diagrams contain charged-lepton propagators and therefore in Table\,\ref{table:WW} we present the contributions to $F_{WW}$ from each of three lepton flavors, $\ell=e,\mu,\tau$, separately.

\begin{table}[t]
\centering
\begin{tabular}{c|c|cc|cc}
\hline\hline
$(s,\Delta)$ & $F_{WW}$ & Type 1 & $|\ell^+\nu\rangle$ \& $|K^+\pi^+\ell^-\bar{\nu}\rangle$ & Type 2 & $|\pi^0\ell^+\nu\rangle$ \\
\hline
\multirow{3}{*}{$(0,0)$} & $e$& $-2.092(50)\times10^{-2}$ & $-2.238(6)\times10^{-2}$ & $1.221(38)\times10^{-1}$ & $-$\\
                         & $\mu$ & $-2.374(47)\times10^{-2}$ & $-2.437(7)\times10^{-2}$ & $1.195(38)\times10^{-1}$ & $-0.504(3)\times10^{-3}$\\
    & $\tau$ & $0.820(79)\times10^{-3}$ & $1.009(6)\times10^{-3}$ & $3.86(14)\times10^{-2}$ & $7.51(4)\times10^{-5}$ \\
\hline
\multirow{3}{*}{$(0,\Delta_{\mathrm{max}}/2)$} & $e$& $-2.122(47)\times10^{-2}$ & $-2.238(6)\times10^{-2}$ & $1.199(38)\times10^{-1}$ & $-$\\
                                & $\mu$ & $-2.528(38)\times10^{-2}$ & $-2.586(7)\times10^{-2}$ & $1.185(37)\times10^{-1}$ & $0.505(8)\times10^{-3}$\\
                                & $\tau$ & $0.455(77)\times10^{-3}$ & $0.603(2)\times10^{-3}$ & $3.84(14)\times10^{-2}$ & $6.93(3)\times10^{-5}$ \\
\hline
\multirow{3}{*}{$(s_{\mathrm{max}}/2,0)$} & $e$& $-2.048(65)\times10^{-2}$ & $-2.238(6)\times10^{-2}$ & $1.225(43)\times10^{-1}$ & $-$\\
                                & $\mu$ & $-2.393(59)\times10^{-2}$ & $-2.489(7)\times10^{-2}$ & $1.190(43)\times10^{-1}$ & $-1.532(17)\times10^{-3}$\\
                                & $\tau$ & $0.535(92)\times10^{-3}$ & $0.811(2)\times10^{-3}$ & $3.93(16)\times10^{-2}$ & $7.43(3)\times10^{-5}$ \\
\hline
\multirow{3}{*}{$(s_{\mathrm{max}}/3,\Delta_{\mathrm{max}}/3)$} & $e$& $-2.103(55)\times10^{-2}$ & $-2.238(6)\times10^{-2}$ & $1.169(42)\times10^{-1}$ & $-$\\
                                & $\mu$ & $-2.525(48)\times10^{-2}$ & $-2.583(7)\times10^{-2}$ & $1.155(41)\times10^{-1}$ & $0.534(6)\times10^{-3}$\\
                                & $\tau$ & $0.354(87)\times10^{-3}$ & $0.608(2)\times10^{-3}$ & $3.74(15)\times10^{-2}$ & $7.06(3)\times10^{-5}$ \\
\hline
\end{tabular}
\caption{Lattice results for the scalar amplitude $F_{WW}(s,\Delta)$ from the $W$-$W$ diagrams.  We also show the results from the lowest intermediate states for each type of diagram. All the results are presented in lattice units.}
\label{table:WW}
\end{table}

   \begin{figure}
    \centering
\includegraphics[width=0.8\textwidth]{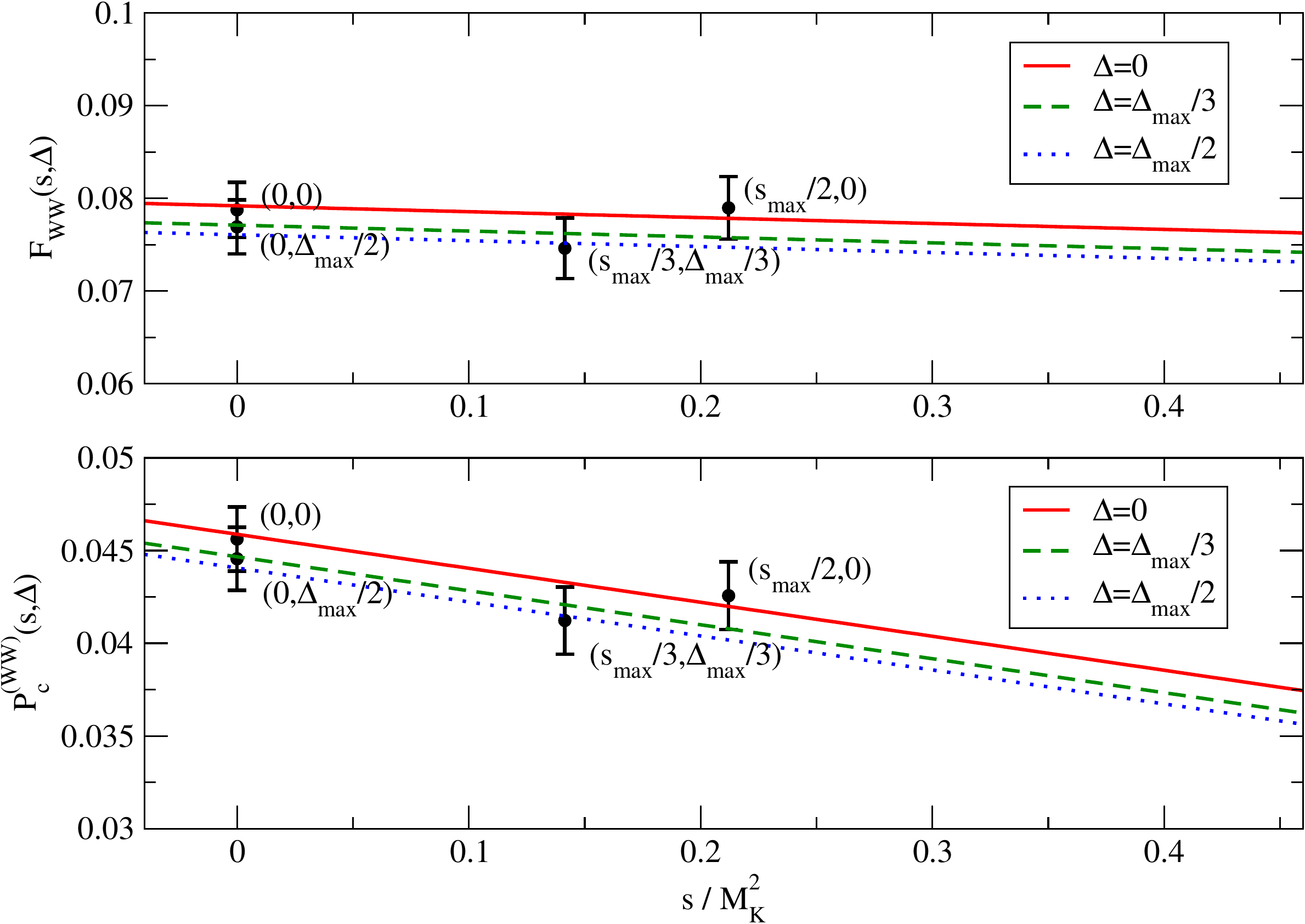}
   \caption{$F_{WW}$ (top panel) and $P_c^{(WW)}$ (bottom panel) as functions of $s$ and $\Delta$.}
    \label{fig:mom_dep_WW}
    \end{figure}

Combining the contributions from Type 1 and Type 2 diagrams and from the three lepton flavors, we obtain the results for $F_{WW}$. In the upper panel of Fig.\,\ref{fig:mom_dep_WW}, we show the $s$ and $\Delta$ dependence of $F_{WW}(s,\Delta)$. In the lower panel, we present the results for $P_c^{(WW)}(s,\Delta)$, which is obtained from $F_{WW}$ using
\ba
P_c^{(WW)}(s,\Delta)=\frac{\pi^2}{\lambda^4M_W^2}\frac{F_{WW}(s,\Delta)}{2f_+(s)}\,.
\ea 
We parametrize the momentum dependence of $P_c^{(WW)}(s,\Delta)$ using the linear function
\ba
P_c^{(WW)}(s,\Delta)=P_c^{(WW)}(0,0)+b_s^{(WW)}\frac{s}{m_K^2}+b_\Delta^{(WW)}\frac{\Delta}{m_K^2}
\ea
and from the correlated fit to the lattice data, we find $b_s^{(WW)}=-(1.8\pm 0.9)\times 10^{-2}$ and $b_\Delta^{(WW)}=-(4.1\pm 0.7)\times10^{-3}$.

As estimated in Ref.\,\cite{Bai:2018hqu}, using the linear parametrization of $P_c(s,\Delta)$, the branching ratio of $K^+\to\pi^+\nu\bar{\nu}$ is proportional to
\ba
\mathrm{Br}(K^+\to\pi^+\nu\bar{\nu})\propto 1+0.071\,b_\Delta^2+0.202\, b_s\,,
\ea
where $b_\Delta=b_\Delta^{(WW)}$ and $b_s=b_s^{(Z)}+b_s^{(WW)}$. Using our determination of $b_\Delta$ and $b_s$ as input, the $s$- and $\Delta$-dependence of $P_c^{(WW)}(s,\Delta)$ only affects the branching ratio at the negligible sub-percent level.

The observation that the momentum dependence is so mild provides a useful guide for our future lattice computations of the $K^+\to\pi^+\nu\bar{\nu}$ decay amplitude. To perform the calculation at physical kinematics is very challenging: on the one hand one needs a large volume in order to accommodate a pion with a mass of about 140\,MeV and on the other hand one needs a fine lattice to avoid lattice artifacts from the physical mass of the charm quark. 
We would require very significant computational resources in order to perform the calculations for a wide range of values of $(s,\Delta)$. 
The study reported here suggests that the $s$- and $\Delta$-dependence of the scalar amplitude has only a minimal effect on the branching ratio. 
Thus even computing the $K^+\to\pi^+\nu\bar{\nu}$ decay amplitude at a single kinematic point $(s,\Delta)$, we can obtain a good estimate of 
the LD contribution to the decay rate.

\section{Contribution of the two-pion intermediate state}\label{sec:pipi}

   \begin{figure}
   \centering
\includegraphics[width=0.8\textwidth]{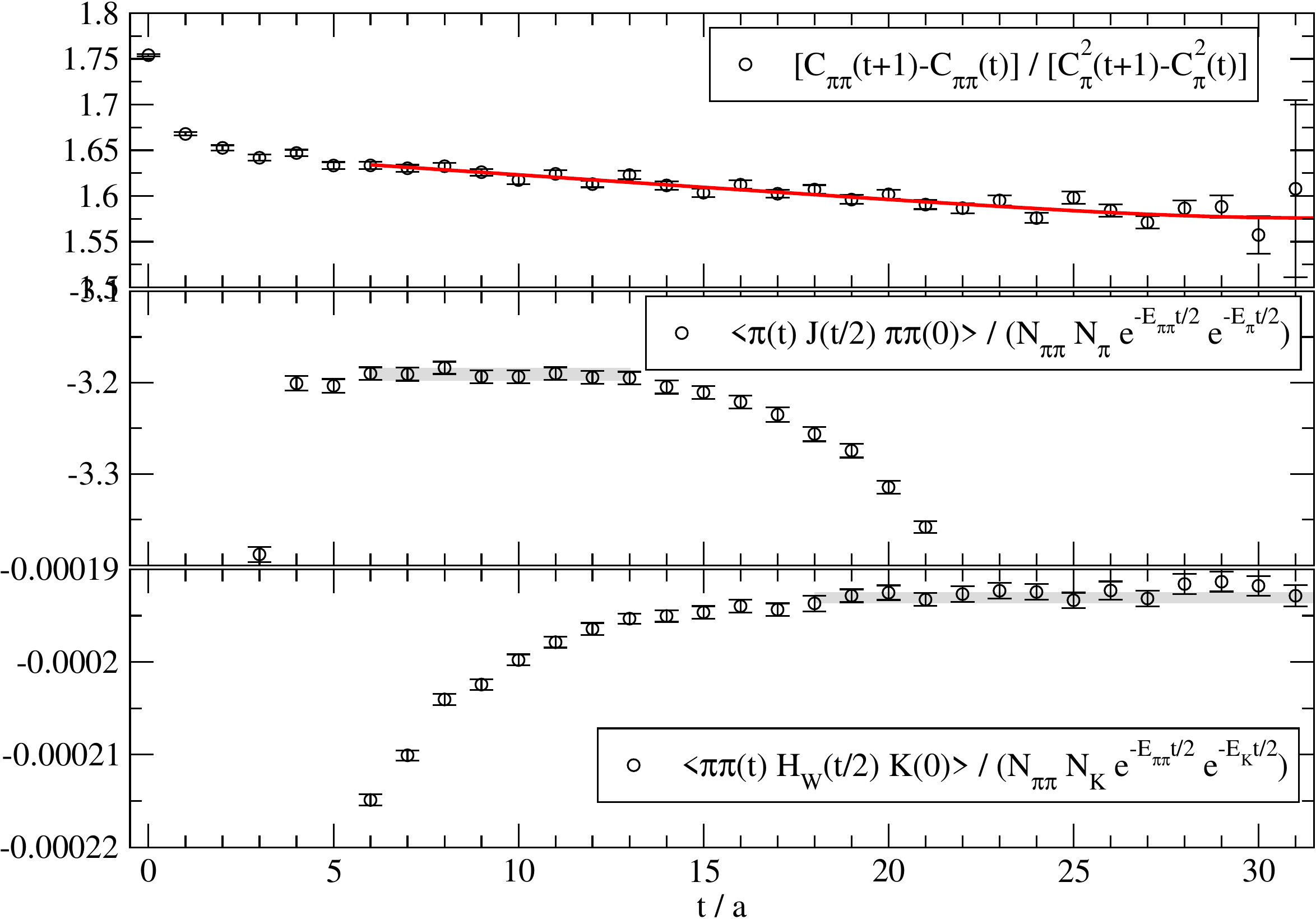}
  \caption{The time dependence of each of the quantities indicated in the boxes.}
  \label{fig:pipi}
  \end{figure}

Two of the less familiar obstacles that must be overcome when computing the $K^+\to\pi^+\nu\bar{\nu}$ decay 
amplitude using lattice QCD are the potentially large finite-volume distortions of the two-pion intermediate state energy spectrum at low energies and the unphysical terms which grow 
exponentially as the Euclidean time separation between the initial- and final-state interpolating 
operators is increased. Such terms which arise from intermediate two-pion states with energy $E_{\pi\pi}$ below $M_K$.   In this section we discuss the corrections that we make for these effects when $s=s_{\mathrm{max}}$.  Thus, for $K^+ \to (\pi^+\pi^0)^{I=2} \to \pi^+\nu\bar\nu$ we need to study contributions from 
the $|\pi^+\pi^0\rangle$ intermediate state with $I = 2$.  We first determine its energy.
In the top panel of Fig.\,\ref{fig:pipi} we show the ratio
\ba\label{eq:Rt}
R(t)=\frac{C_{\pi\pi}(t+1)-C_{\pi\pi}(t)}{C_\pi^2(t+1)-C^2_\pi(t)}
\ea
as a function of $t$. Here $C_{\pi\pi}(t)$ is a two-point correlation function for the $(\pi\pi)^{I=2}$ interpolating operator in the center-of-mass frame and $C_\pi(t)$ is a correlation function for the single $\pi$ interpolating operator at rest (see Ref.\,\cite{Bai:2018hqu} for the choice of interpolating operators). The subtraction in 
$C_{\pi\pi}(t+1)-C_{\pi\pi}(t)$
is introduced to remove the constant term arising 
from the around-the-world effect and the corresponding difference $C_\pi^2(t+1)-C^2_\pi(t)$ is
used in the denominator to insure that the ratio $R(t)$ in Eq.\,(\ref{eq:Rt}) is proportional to  
$\exp(-\Delta E)$.  Thus, by studying the $t$-dependence of $R(t)$ as explained in Ref.\,\cite{Feng:2009ij}, we determine the energy difference $\Delta E=E_{\pi\pi}-2m_\pi=0.001687(41)$ (note that the pion mass in lattice units is 0.12535(27)). Using L\"uscher's finite-volume formula, we find the $(\pi\pi)^{I=2}$ scattering length to be given by 
$m_\pi a_{\pi\pi}^{I=2}=-0.0659(16)$, in good agreement with the prediction from leading-order chiral perturbation theory, $m_\pi a_{\pi\pi}^{\mathrm{LO}}=-\frac{m_\pi^2}{8\pi f_\pi^2}=-0.0673(5)$.

   \begin{figure}
   \centering
\includegraphics[width=0.8\textwidth]{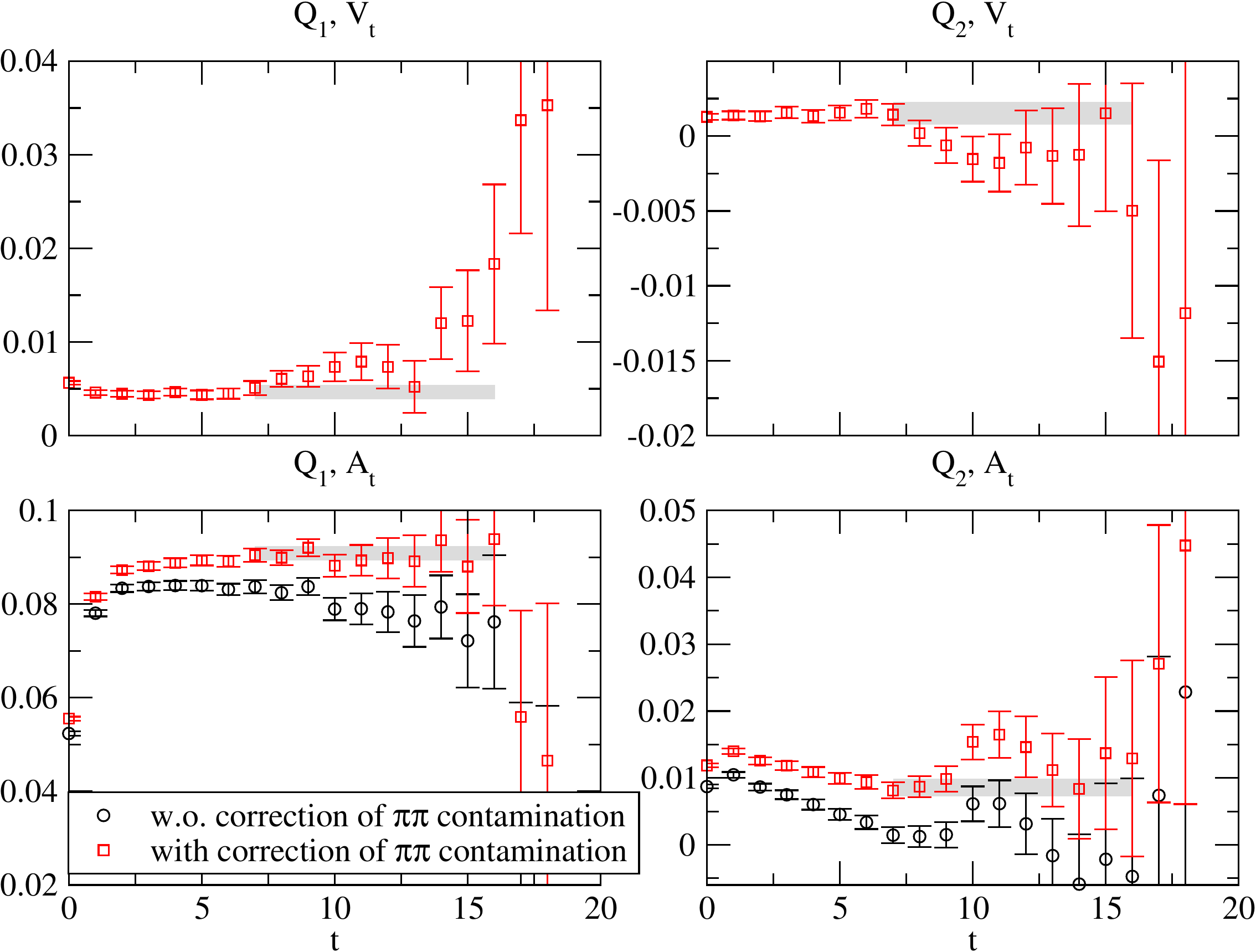}
  \caption{Contributions to $F_0^{Z,\pi\pi}(s_\mathrm{max})$ from the vector (top row) and axial vector (bottom row) components of the quark currents for the $\Delta S=1$ operators $Q_1$ (left column) and $Q_2$ (right column). With the axial current, there are unphysical exponentially growing contributions from the two-pion intermediate state and the results are shown with and without the subtraction of these terms.}
  \label{fig:pipi_contamination}
  \end{figure}

In the next step, we calculate the $_{\,0}\langle\pi\pi|H_W|K\rangle$ and $\langle\pi|J_\mu|\pi\pi\rangle_0$ matrix elements, where the subscript $0$ indicates the finite-volume two-pion ground state, and construct the two ratios:
\begin{equation}
R_{\pi\pi\to\pi}(t)=\frac{\langle\pi(t)J_\mu(t/2)\pi\pi(0)\rangle}{N_{\pi\pi}N_\pi e^{-E_{\pi\pi}t/2} e^{-E_\pi t/2}}\,,\quad
R_{K\to\pi\pi}(t)=\frac{\langle\pi\pi(t)H_W(t/2)K(0)\rangle}{N_{\pi\pi}N_K e^{-E_{\pi\pi}t/2} e^{-E_K t/2}}\,,
\end{equation}
where the normalisation factors are $N_\pi=\langle 0\,|\pi(0)|\pi(\vec 0\,)\rangle$,  
$N_K=\langle 0\,|K(0)|K(\vec 0\,)\rangle$ and $N_{\pi\pi}=\langle 0\,|\pi\pi(0)|\pi\pi\rangle_0$.
We present $R_{\pi\pi\to\pi}(t)$ and $R_{K\to\pi\pi}(t)$ as functions of $t$ in the central and lower panels of Fig.\,\ref{fig:pipi} respectively.
Performing fits for $R_{\pi\pi\to\pi}(t)$ and $R_{K\to\pi\pi}(t)$ in the plateau regions, as indicated by the gray bands in Fig.\,\ref{fig:pipi},
we determine the matrix elements to be  $_0\langle\pi\pi|H_W|K\rangle=-i\,9.653(25)\times10^{-5}$ and $\langle\pi|J_4|\pi\pi\rangle_0=i\,3.1910(60)$. The value of $\langle\pi|J_4|\pi\pi\rangle_0$ is close to the factorization approximation 
$\langle\pi|J_4|\pi\pi\rangle_0\approx \langle\pi|\pi\rangle \langle0|J_4|\pi\rangle=i\,3.2149(62)$. Using these 
matrix elements as inputs, we find that the $\pi\pi$ contribution to $F_0^Z(s_{\mathrm{max}})$ is $F_0^{Z,\pi\pi}(s_{\mathrm{max}})=-1.536(5)\times10^{-3}$, which is about 7.5\% of the total scalar amplitude $F_0^Z(s_{\mathrm{max}})$. Such effects are not small and result in a sizeable exponentially growing contamination. In Fig.\,\ref{fig:pipi_contamination}, we show the lattice results with and without correcting for the exponentially growing contamination from the $\pi\pi$ intermediate state. The corresponding systematic effects are significant when compared to the statistical ones. We thus include these corrections in our analysis when producing the results shown in Table\,\ref{tab:lattice_results_Z_exchange}.

In addition to removing the exponentially growing contamination, we also evaluate and correct for the corresponding finite-volume effects due to the $\pi\pi$ intermediate state. As shown in Ref.\,\cite{Bai:2018hqu,Christ:2015pwa}, 
the finite-volume correction which must be subtracted from the
finite-volume amplitude to obtain the infinite-volume result is
\ba\label{eq:DeltaFV}
\Delta_{\mathrm{FV}}=\frac{d\big(\phi(E_{\pi\pi})+\delta(E_{\pi\pi})\big)}{dE_{\pi\pi}}\cot(\phi+\delta)\langle \pi|J_\mu|\pi\pi\rangle\langle\pi\pi|H_W|K\rangle\,,
\ea
where $\phi$ is a kinematic function of $E_{\pi\pi}$ and $\delta(E_{\pi\pi})$ is the $I=2$ $s$-wave $\pi\pi$ phase shift.
Note that in Eq.\,(\ref{eq:DeltaFV}) all quantities are to be evaluated at $E_{\pi\pi} = M_K$
and the two-pion matrix elements must be evaluated in a spatial volume 
that has been chosen so that the finite-volume state $|\pi\pi\rangle$ has 
the energy of the kaon. 
 We use the approximation $\delta\approx k\, a_{\pi\pi}$
 where $k=\sqrt{E_{\pi\pi}^2/4-m_\pi^2}$ and $d\delta/dE_{\pi\pi}\approx a_{\pi\pi}\,\frac{E_{\pi\pi}}{4k}$ to estimate $\Delta_{\mathrm{FV}}$ and find $\Delta_{FV}=4.28(7)\times10^{-4}$, corresponding to a correction to $P_c$ of about $4.8\times 10^{-4}$. Here we approximate
the energy-conserving matrix elements needed in Eq. (\ref{eq:DeltaFV}) by those
determined above when $E_{\pi\pi} = 0.25239(60)$. Since the expected value of the full charm-quark contribution to $P_c$ is about 0.4\,\cite{Buras:2015qea}, this finite-volume correction has a negligible, per-mille effect on the branching ratio (see Eq.\,(97) and the following discussion in Ref.\,\cite{Bai:2018hqu}).

\section{Summary and Conclusions}\label{sec:concs}
In this paper we report on a study of the long-distance contribution to the amplitude of the rare kaon decay $K^+\to\pi^+\nu\bar\nu$ at the near physical pion mass of $m_\pi=170$\,MeV. 
This study is an intermediate step in our long-term project to evaluate the rate for this decay at physical kinematics and with sub-percent precision. 
Having previously developed the necessary theoretical framework for the evaluation of the long-distance contributions in Ref.\,\cite{Christ:2016eae} and performed a complete computation at unphysical quark masses and for a single choice of meson and neutrino momenta in Refs.\,\cite{Bai:2017fkh,Bai:2018hqu}, in this paper we investigate two areas to guide us and to provide experience for our future calculations at physical masses. 
These are:\\[0.1cm]
{1.} \emph{The momentum dependence of the amplitude.} Since varying the momenta of the mesons and neutrinos requires very significant computational resources, it was instructive to gain some understanding of the behaviour of the amplitude as a function of the momenta.  Given the limited kinematic range available, we expect the momentum dependence to be very mild  and in Sec.\,\ref{sec:results} we explicitly confirm this and conclude that a computation at physical masses even at a single kinematic point $(s,\Delta)$ would provide a good estimate of the long-distance contribution to the decay rate. Of course, eventually as computing resources allow, this conclusion can be tested also at physical quark masses. 

An additional conclusion from this study is that the contribution from disconnected diagrams appears to be of the order of 2\% of that coming from the connected part. This was the case at $s=s_\mathrm{max}$, which is the only point at which disconnected diagrams could be evaluated, since it was at this point that twisted boundary conditions for the quarks were not used.\\[0.1cm]
2. \emph{Contribution from $\pi\pi$ intermediate states:} With meson masses $m_K=493(3)$\,MeV and $m_\pi=172(1)$\,MeV it is possible to have isospin 2 $\pi\pi$ intermediate states with energy below $M_K$, $K^+\to(\pi^+\pi^0)^{I=2}\to\pi^+\nu\bar\nu$, from the $Z$-exchange diagrams. Two consequence of this in the evaluation of long-distance contributions in a finite-volume Euclidean-space calculation are:\\
(i) the presence of unphysical terms which grow exponentially with the range of the time-separation of the two weak operators and which must be subtracted;\\
(ii) finite-volume effects which are not exponentially small in the volume and which must be removed.\\
The procedures to remove the exponentially growing terms and the power-like finite-volume corrections are well understood. In Sec.\,\ref{sec:pipi} we perform the subtraction of the exponentially growing terms and find that 
the contribution from the two-pion intermediate state is a little less than 1\%
of the full decay amplitude.  We 
calculate the corresponding FV effects and find them to be negligibly small. 

As the final step in this $K^+\to\pi^+\nu\bar\nu$ project we have generated data on a
$64^3\times128$ lattice with M\"obius domain wall fermions and the Iwasaki gauge
action at an inverse lattice spacing of $a^{-1}=2.359(7)$\,GeV, pion mass
$m_\pi=135.9(3)$\,MeV and kaon mass $m_K=496.9(7)$\,MeV~\cite{Blum:2014tka}.
We use a series of charm quark masses surrounding the physical value.
The use of this ensemble will remove the significant systematic effect arising from the lighter than-physical charm quark mass of the present study, resulting in a physical rare kaon decay amplitude with the main systematic effects under control.
The resources required for the $64^3\times128$ calculation are relatively demanding making the determination of the amplitude at several kinematic points computationally expensive. 
The observation of the very mild momentum dependence of the decay
amplitude observed in this study implies that we can obtain the rate with good precision, even if we perform a computation at a single
kinematic point of $(s,\Delta)$.

\vspace{-0.2in}
\section*{Acknowledgements}

\vspace{-0.2in}
X.F. was supported in part by NSFC of China under Grant No. 11775002.
C.T.S.~was partially supported by STFC (UK) grant No. ST/P000711/1 and by an Emeritus Fellowship from the Leverhulme Trust.
A.P. is supported in part by UK STFC grants ST/P000630/1. A.P. also received
funding from the European Research Council (ERC) under the European Union's
Horizon 2020 research and innovation programme under grant agreements No 757646
\& 813942.

\vfill
\bibliography{paper}

\begin{thebibliography}{34}%
\makeatletter
\providecommand \@ifxundefined [1]{%
 \@ifx{#1\undefined}
}%
\providecommand \@ifnum [1]{%
 \ifnum #1\expandafter \@firstoftwo
 \else \expandafter \@secondoftwo
 \fi
}%
\providecommand \@ifx [1]{%
 \ifx #1\expandafter \@firstoftwo
 \else \expandafter \@secondoftwo
 \fi
}%
\providecommand \natexlab [1]{#1}%
\providecommand \enquote  [1]{``#1''}%
\providecommand \bibnamefont  [1]{#1}%
\providecommand \bibfnamefont [1]{#1}%
\providecommand \citenamefont [1]{#1}%
\providecommand \href@noop [0]{\@secondoftwo}%
\providecommand \href [0]{\begingroup \@sanitize@url \@href}%
\providecommand \@href[1]{\@@startlink{#1}\@@href}%
\providecommand \@@href[1]{\endgroup#1\@@endlink}%
\providecommand \@sanitize@url [0]{\catcode `\\12\catcode `\$12\catcode
  `\&12\catcode `\#12\catcode `\^12\catcode `\_12\catcode `\%12\relax}%
\providecommand \@@startlink[1]{}%
\providecommand \@@endlink[0]{}%
\providecommand \url  [0]{\begingroup\@sanitize@url \@url }%
\providecommand \@url [1]{\endgroup\@href {#1}{\urlprefix }}%
\providecommand \urlprefix  [0]{URL }%
\providecommand \Eprint [0]{\href }%
\providecommand \doibase [0]{http://dx.doi.org/}%
\providecommand \selectlanguage [0]{\@gobble}%
\providecommand \bibinfo  [0]{\@secondoftwo}%
\providecommand \bibfield  [0]{\@secondoftwo}%
\providecommand \translation [1]{[#1]}%
\providecommand \BibitemOpen [0]{}%
\providecommand \bibitemStop [0]{}%
\providecommand \bibitemNoStop [0]{.\EOS\space}%
\providecommand \EOS [0]{\spacefactor3000\relax}%
\providecommand \BibitemShut  [1]{\csname bibitem#1\endcsname}%
\let\auto@bib@innerbib\@empty
\bibitem [{\citenamefont {Isidori}\ \emph {et~al.}(2005)\citenamefont
  {Isidori}, \citenamefont {Mescia},\ and\ \citenamefont
  {Smith}}]{Isidori:2005xm}%
  \BibitemOpen
  \bibfield  {author} {\bibinfo {author} {\bibfnamefont {G.}~\bibnamefont
  {Isidori}}, \bibinfo {author} {\bibfnamefont {F.}~\bibnamefont {Mescia}}, \
  and\ \bibinfo {author} {\bibfnamefont {C.}~\bibnamefont {Smith}},\ }\href
  {\doibase 10.1016/j.nuclphysb.2005.04.008} {\bibfield  {journal} {\bibinfo
  {journal} {Nucl.Phys.}\ }\textbf {\bibinfo {volume} {B718}},\ \bibinfo
  {pages} {319} (\bibinfo {year} {2005})},\ \Eprint
  {http://arxiv.org/abs/hep-ph/0503107} {arXiv:hep-ph/0503107 [hep-ph]}
  \BibitemShut {NoStop}%
\bibitem [{\citenamefont {Buras}\ \emph {et~al.}(2015)\citenamefont {Buras},
  \citenamefont {Buttazzo}, \citenamefont {Girrbach-Noe},\ and\ \citenamefont
  {Knegjens}}]{Buras:2015qea}%
  \BibitemOpen
  \bibfield  {author} {\bibinfo {author} {\bibfnamefont {A.~J.}\ \bibnamefont
  {Buras}}, \bibinfo {author} {\bibfnamefont {D.}~\bibnamefont {Buttazzo}},
  \bibinfo {author} {\bibfnamefont {J.}~\bibnamefont {Girrbach-Noe}}, \ and\
  \bibinfo {author} {\bibfnamefont {R.}~\bibnamefont {Knegjens}},\ }\href
  {\doibase 10.1007/JHEP11(2015)033} {\bibfield  {journal} {\bibinfo  {journal}
  {JHEP}\ }\textbf {\bibinfo {volume} {11}},\ \bibinfo {pages} {033} (\bibinfo
  {year} {2015})},\ \Eprint {http://arxiv.org/abs/1503.02693} {arXiv:1503.02693
  [hep-ph]} \BibitemShut {NoStop}%
\bibitem [{\citenamefont {Artamonov}\ \emph {et~al.}(2008)\citenamefont
  {Artamonov} \emph {et~al.}}]{Artamonov:2008qb}%
  \BibitemOpen
  \bibfield  {author} {\bibinfo {author} {\bibfnamefont {A.}~\bibnamefont
  {Artamonov}} \emph {et~al.} (\bibinfo {collaboration} {E949 Collaboration}),\
  }\href {\doibase 10.1103/PhysRevLett.101.191802} {\bibfield  {journal}
  {\bibinfo  {journal} {Phys.Rev.Lett.}\ }\textbf {\bibinfo {volume} {101}},\
  \bibinfo {pages} {191802} (\bibinfo {year} {2008})},\ \Eprint
  {http://arxiv.org/abs/0808.2459} {arXiv:0808.2459 [hep-ex]} \BibitemShut
  {NoStop}%
\bibitem [{\citenamefont {Adler}\ \emph {et~al.}(1997)\citenamefont {Adler}
  \emph {et~al.}}]{Adler:1997am}%
  \BibitemOpen
  \bibfield  {author} {\bibinfo {author} {\bibfnamefont {S.}~\bibnamefont
  {Adler}} \emph {et~al.} (\bibinfo {collaboration} {E787 Collaboration}),\
  }\href {\doibase 10.1103/PhysRevLett.79.2204} {\bibfield  {journal} {\bibinfo
   {journal} {Phys.Rev.Lett.}\ }\textbf {\bibinfo {volume} {79}},\ \bibinfo
  {pages} {2204} (\bibinfo {year} {1997})},\ \Eprint
  {http://arxiv.org/abs/hep-ex/9708031} {arXiv:hep-ex/9708031 [hep-ex]}
  \BibitemShut {NoStop}%
\bibitem [{\citenamefont {Adler}\ \emph {et~al.}(2000)\citenamefont {Adler}
  \emph {et~al.}}]{Adler:2000by}%
  \BibitemOpen
  \bibfield  {author} {\bibinfo {author} {\bibfnamefont {S.}~\bibnamefont
  {Adler}} \emph {et~al.} (\bibinfo {collaboration} {E787 Collaboration}),\
  }\href {\doibase 10.1103/PhysRevLett.84.3768} {\bibfield  {journal} {\bibinfo
   {journal} {Phys.Rev.Lett.}\ }\textbf {\bibinfo {volume} {84}},\ \bibinfo
  {pages} {3768} (\bibinfo {year} {2000})},\ \Eprint
  {http://arxiv.org/abs/hep-ex/0002015} {arXiv:hep-ex/0002015 [hep-ex]}
  \BibitemShut {NoStop}%
\bibitem [{\citenamefont {Adler}\ \emph
  {et~al.}(2002{\natexlab{a}})\citenamefont {Adler} \emph
  {et~al.}}]{Adler:2001xv}%
  \BibitemOpen
  \bibfield  {author} {\bibinfo {author} {\bibfnamefont {S.}~\bibnamefont
  {Adler}} \emph {et~al.} (\bibinfo {collaboration} {E787 Collaboration}),\
  }\href {\doibase 10.1103/PhysRevLett.88.041803} {\bibfield  {journal}
  {\bibinfo  {journal} {Phys.Rev.Lett.}\ }\textbf {\bibinfo {volume} {88}},\
  \bibinfo {pages} {041803} (\bibinfo {year} {2002}{\natexlab{a}})},\ \Eprint
  {http://arxiv.org/abs/hep-ex/0111091} {arXiv:hep-ex/0111091 [hep-ex]}
  \BibitemShut {NoStop}%
\bibitem [{\citenamefont {Adler}\ \emph
  {et~al.}(2002{\natexlab{b}})\citenamefont {Adler} \emph
  {et~al.}}]{Adler:2002hy}%
  \BibitemOpen
  \bibfield  {author} {\bibinfo {author} {\bibfnamefont {S.~S.}\ \bibnamefont
  {Adler}} \emph {et~al.} (\bibinfo {collaboration} {E787 Collaboration}),\
  }\href {\doibase 10.1016/S0370-2693(02)01911-1} {\bibfield  {journal}
  {\bibinfo  {journal} {Phys.Lett.}\ }\textbf {\bibinfo {volume} {B537}},\
  \bibinfo {pages} {211} (\bibinfo {year} {2002}{\natexlab{b}})},\ \Eprint
  {http://arxiv.org/abs/hep-ex/0201037} {arXiv:hep-ex/0201037 [hep-ex]}
  \BibitemShut {NoStop}%
\bibitem [{\citenamefont {Anisimovsky}\ \emph {et~al.}(2004)\citenamefont
  {Anisimovsky} \emph {et~al.}}]{Anisimovsky:2004hr}%
  \BibitemOpen
  \bibfield  {author} {\bibinfo {author} {\bibfnamefont {V.}~\bibnamefont
  {Anisimovsky}} \emph {et~al.} (\bibinfo {collaboration} {E949
  Collaboration}),\ }\href {\doibase 10.1103/PhysRevLett.93.031801} {\bibfield
  {journal} {\bibinfo  {journal} {Phys.Rev.Lett.}\ }\textbf {\bibinfo {volume}
  {93}},\ \bibinfo {pages} {031801} (\bibinfo {year} {2004})},\ \Eprint
  {http://arxiv.org/abs/hep-ex/0403036} {arXiv:hep-ex/0403036 [hep-ex]}
  \BibitemShut {NoStop}%
\bibitem [{\citenamefont {Cortina~Gil}\ \emph {et~al.}(2019)\citenamefont
  {Cortina~Gil} \emph {et~al.}}]{CortinaGil:2018fkc}%
  \BibitemOpen
  \bibfield  {author} {\bibinfo {author} {\bibfnamefont {E.}~\bibnamefont
  {Cortina~Gil}} \emph {et~al.} (\bibinfo {collaboration} {NA62}),\ }\href
  {\doibase 10.1016/j.physletb.2019.01.067} {\bibfield  {journal} {\bibinfo
  {journal} {Phys. Lett.}\ }\textbf {\bibinfo {volume} {B791}},\ \bibinfo
  {pages} {156} (\bibinfo {year} {2019})},\ \Eprint
  {http://arxiv.org/abs/1811.08508} {arXiv:1811.08508 [hep-ex]} \BibitemShut
  {NoStop}%
\bibitem [{\citenamefont {Yamanaka}(2012)}]{Yamanaka:2012yma}%
  \BibitemOpen
  \bibfield  {author} {\bibinfo {author} {\bibfnamefont {T.}~\bibnamefont
  {Yamanaka}} (\bibinfo {collaboration} {KOTO}),\ }\href {\doibase
  10.1093/ptep/pts057} {\bibfield  {journal} {\bibinfo  {journal} {PTEP}\
  }\textbf {\bibinfo {volume} {2012}},\ \bibinfo {pages} {02B006} (\bibinfo
  {year} {2012})}\BibitemShut {NoStop}%
\bibitem [{\citenamefont {Ahn}\ \emph {et~al.}(2017)\citenamefont {Ahn} \emph
  {et~al.}}]{Ahn:2016kja}%
  \BibitemOpen
  \bibfield  {author} {\bibinfo {author} {\bibfnamefont {J.~K.}\ \bibnamefont
  {Ahn}} \emph {et~al.} (\bibinfo {collaboration} {KOTO}),\ }\href {\doibase
  10.1093/ptep/ptx001} {\bibfield  {journal} {\bibinfo  {journal} {PTEP}\
  }\textbf {\bibinfo {volume} {2017}},\ \bibinfo {pages} {021C01} (\bibinfo
  {year} {2017})},\ \Eprint {http://arxiv.org/abs/1609.03637} {arXiv:1609.03637
  [hep-ex]} \BibitemShut {NoStop}%
\bibitem [{\citenamefont {Shinohara}()}]{Shinohara:2019xxx}%
  \BibitemOpen
  \bibfield  {author} {\bibinfo {author} {\bibfnamefont {S.}~\bibnamefont
  {Shinohara}},\ }\href
  {https://indico.cern.ch/event/769729/contributions/3510939/attachments/1904988/3145907/KAON2019_shinohara_upload.pdf}
  {\enquote {\bibinfo {title} {{Search for the rare decay
  $K_L\to\pi^0\nu\bar\nu$ at J-PARC KOTO Experiment}},}\ }\bibinfo {note}
  {{Presented at the International Conference on Kaon Physics 2019
  (Kaon2019)}}\BibitemShut {NoStop}%
\bibitem [{\citenamefont {Sachrajda}(2013{\natexlab{a}})}]{Sachrajda:2013vqa}%
  \BibitemOpen
  \bibfield  {author} {\bibinfo {author} {\bibfnamefont {C.~T.}\ \bibnamefont
  {Sachrajda}} (\bibinfo {collaboration} {RBC-UKQCD}),\ }\bibfield  {booktitle}
  {\emph {\bibinfo {booktitle} {{Proceedings, 7th International Workshop on
  Chiral Dynamics (CD12): Newport News, Virginia, USA, August 6-10, 2012}}},\
  }\href@noop {} {\bibfield  {journal} {\bibinfo  {journal} {PoS}\ }\textbf
  {\bibinfo {volume} {CD12}},\ \bibinfo {pages} {009} (\bibinfo {year}
  {2013}{\natexlab{a}})}\BibitemShut {NoStop}%
\bibitem [{\citenamefont {Sachrajda}(2013{\natexlab{b}})}]{Sachrajda:2013fxa}%
  \BibitemOpen
  \bibfield  {author} {\bibinfo {author} {\bibfnamefont {C.~T.}\ \bibnamefont
  {Sachrajda}} (\bibinfo {collaboration} {RBC-UKQCD}),\ }\bibfield  {booktitle}
  {\emph {\bibinfo {booktitle} {{Proceedings, Kaon Physics International
  Conference (KAON13): Ann Arbor, Michigan, April 29-May 1, 2013}}},\
  }\href@noop {} {\bibfield  {journal} {\bibinfo  {journal} {PoS}\ }\textbf
  {\bibinfo {volume} {KAON13}},\ \bibinfo {pages} {019} (\bibinfo {year}
  {2013}{\natexlab{b}})}\BibitemShut {NoStop}%
\bibitem [{\citenamefont {Feng}\ \emph {et~al.}(2015)\citenamefont {Feng},
  \citenamefont {Christ}, \citenamefont {Portelli},\ and\ \citenamefont
  {Sachrajda}}]{Feng:2015kfa}%
  \BibitemOpen
  \bibfield  {author} {\bibinfo {author} {\bibfnamefont {X.}~\bibnamefont
  {Feng}}, \bibinfo {author} {\bibfnamefont {N.~H.}\ \bibnamefont {Christ}},
  \bibinfo {author} {\bibfnamefont {A.}~\bibnamefont {Portelli}}, \ and\
  \bibinfo {author} {\bibfnamefont {C.}~\bibnamefont {Sachrajda}},\ }\bibfield
  {booktitle} {\emph {\bibinfo {booktitle} {{Proceedings, 32nd International
  Symposium on Lattice Field Theory (Lattice 2014): Brookhaven, NY, USA, June
  23-28, 2014}}},\ }\href@noop {} {\bibfield  {journal} {\bibinfo  {journal}
  {PoS}\ }\textbf {\bibinfo {volume} {LATTICE2014}},\ \bibinfo {pages} {367}
  (\bibinfo {year} {2015})}\BibitemShut {NoStop}%
\bibitem [{\citenamefont {Christ}\ \emph
  {et~al.}(2015{\natexlab{a}})\citenamefont {Christ}, \citenamefont {Feng},
  \citenamefont {Portelli},\ and\ \citenamefont {Sachrajda}}]{Christ:2015aha}%
  \BibitemOpen
  \bibfield  {author} {\bibinfo {author} {\bibfnamefont {N.~H.}\ \bibnamefont
  {Christ}}, \bibinfo {author} {\bibfnamefont {X.}~\bibnamefont {Feng}},
  \bibinfo {author} {\bibfnamefont {A.}~\bibnamefont {Portelli}}, \ and\
  \bibinfo {author} {\bibfnamefont {C.~T.}\ \bibnamefont {Sachrajda}} (\bibinfo
  {collaboration} {RBC, UKQCD}),\ }\href {\doibase 10.1103/PhysRevD.92.094512}
  {\bibfield  {journal} {\bibinfo  {journal} {Phys. Rev.}\ }\textbf {\bibinfo
  {volume} {D92}},\ \bibinfo {pages} {094512} (\bibinfo {year}
  {2015}{\natexlab{a}})},\ \Eprint {http://arxiv.org/abs/1507.03094}
  {arXiv:1507.03094 [hep-lat]} \BibitemShut {NoStop}%
\bibitem [{\citenamefont {Christ}\ \emph
  {et~al.}(2016{\natexlab{a}})\citenamefont {Christ}, \citenamefont {Feng},
  \citenamefont {Portelli},\ and\ \citenamefont {Sachrajda}}]{Christ:2016eae}%
  \BibitemOpen
  \bibfield  {author} {\bibinfo {author} {\bibfnamefont {N.~H.}\ \bibnamefont
  {Christ}}, \bibinfo {author} {\bibfnamefont {X.}~\bibnamefont {Feng}},
  \bibinfo {author} {\bibfnamefont {A.}~\bibnamefont {Portelli}}, \ and\
  \bibinfo {author} {\bibfnamefont {C.~T.}\ \bibnamefont {Sachrajda}} (\bibinfo
  {collaboration} {RBC, UKQCD}),\ }\href {\doibase 10.1103/PhysRevD.93.114517}
  {\bibfield  {journal} {\bibinfo  {journal} {Phys. Rev.}\ }\textbf {\bibinfo
  {volume} {D93}},\ \bibinfo {pages} {114517} (\bibinfo {year}
  {2016}{\natexlab{a}})},\ \Eprint {http://arxiv.org/abs/1605.04442}
  {arXiv:1605.04442 [hep-lat]} \BibitemShut {NoStop}%
\bibitem [{\citenamefont {Christ}\ \emph
  {et~al.}(2016{\natexlab{b}})\citenamefont {Christ}, \citenamefont {Feng},
  \citenamefont {Jüttner}, \citenamefont {Lawson}, \citenamefont {Portelli},\
  and\ \citenamefont {Sachrajda}}]{Christ:2016psm}%
  \BibitemOpen
  \bibfield  {author} {\bibinfo {author} {\bibfnamefont {N.~H.}\ \bibnamefont
  {Christ}}, \bibinfo {author} {\bibfnamefont {X.}~\bibnamefont {Feng}},
  \bibinfo {author} {\bibfnamefont {A.}~\bibnamefont {Jüttner}}, \bibinfo
  {author} {\bibfnamefont {A.}~\bibnamefont {Lawson}}, \bibinfo {author}
  {\bibfnamefont {A.}~\bibnamefont {Portelli}}, \ and\ \bibinfo {author}
  {\bibfnamefont {C.~T.}\ \bibnamefont {Sachrajda}},\ }\bibfield  {booktitle}
  {\emph {\bibinfo {booktitle} {{Proceedings, 8th International Workshop on
  Chiral Dynamics (CD15): Pisa, Italy, June 29-July 3, 2015}}},\ }\href@noop {}
  {\bibfield  {journal} {\bibinfo  {journal} {PoS}\ }\textbf {\bibinfo {volume}
  {CD15}},\ \bibinfo {pages} {033} (\bibinfo {year}
  {2016}{\natexlab{b}})}\BibitemShut {NoStop}%
\bibitem [{\citenamefont {Christ}\ \emph
  {et~al.}(2016{\natexlab{c}})\citenamefont {Christ}, \citenamefont {Feng},
  \citenamefont {Juttner}, \citenamefont {Lawson}, \citenamefont {Portelli},\
  and\ \citenamefont {Sachrajda}}]{Christ:2016awg}%
  \BibitemOpen
  \bibfield  {author} {\bibinfo {author} {\bibfnamefont {N.}~\bibnamefont
  {Christ}}, \bibinfo {author} {\bibfnamefont {X.}~\bibnamefont {Feng}},
  \bibinfo {author} {\bibfnamefont {A.}~\bibnamefont {Juttner}}, \bibinfo
  {author} {\bibfnamefont {A.}~\bibnamefont {Lawson}}, \bibinfo {author}
  {\bibfnamefont {A.}~\bibnamefont {Portelli}}, \ and\ \bibinfo {author}
  {\bibfnamefont {C.}~\bibnamefont {Sachrajda}},\ }in\ \href
  {http://inspirehep.net/record/1419253/files/arXiv:1602.01374.pdf} {\emph
  {\bibinfo {booktitle} {{Proceedings, 33rd International Symposium on Lattice
  Field Theory (Lattice 2015)}}}}\ (\bibinfo {year} {2016})\ \Eprint
  {http://arxiv.org/abs/1602.01374} {arXiv:1602.01374 [hep-lat]} \BibitemShut
  {NoStop}%
\bibitem [{\citenamefont {Christ}\ \emph
  {et~al.}(2016{\natexlab{d}})\citenamefont {Christ}, \citenamefont {Feng},
  \citenamefont {Lawson}, \citenamefont {Portelli},\ and\ \citenamefont
  {Sachrajda}}]{Christ:2016lro}%
  \BibitemOpen
  \bibfield  {author} {\bibinfo {author} {\bibfnamefont {N.~H.}\ \bibnamefont
  {Christ}}, \bibinfo {author} {\bibfnamefont {X.}~\bibnamefont {Feng}},
  \bibinfo {author} {\bibfnamefont {A.}~\bibnamefont {Lawson}}, \bibinfo
  {author} {\bibfnamefont {A.}~\bibnamefont {Portelli}}, \ and\ \bibinfo
  {author} {\bibfnamefont {C.}~\bibnamefont {Sachrajda}},\ }\bibfield
  {booktitle} {\emph {\bibinfo {booktitle} {{Proceedings, 34th International
  Symposium on Lattice Field Theory (Lattice 2016): Southampton, UK, July
  24-30, 2016}}},\ }\href@noop {} {\bibfield  {journal} {\bibinfo  {journal}
  {PoS}\ }\textbf {\bibinfo {volume} {LATTICE2016}},\ \bibinfo {pages} {306}
  (\bibinfo {year} {2016}{\natexlab{d}})}\BibitemShut {NoStop}%
\bibitem [{\citenamefont {Christ}\ \emph
  {et~al.}(2016{\natexlab{e}})\citenamefont {Christ}, \citenamefont {Feng},
  \citenamefont {Juttner}, \citenamefont {Lawson}, \citenamefont {Portelli},\
  and\ \citenamefont {Sachrajda}}]{Christ:2016mmq}%
  \BibitemOpen
  \bibfield  {author} {\bibinfo {author} {\bibfnamefont {N.~H.}\ \bibnamefont
  {Christ}}, \bibinfo {author} {\bibfnamefont {X.}~\bibnamefont {Feng}},
  \bibinfo {author} {\bibfnamefont {A.}~\bibnamefont {Juttner}}, \bibinfo
  {author} {\bibfnamefont {A.}~\bibnamefont {Lawson}}, \bibinfo {author}
  {\bibfnamefont {A.}~\bibnamefont {Portelli}}, \ and\ \bibinfo {author}
  {\bibfnamefont {C.~T.}\ \bibnamefont {Sachrajda}},\ }\href {\doibase
  10.1103/PhysRevD.94.114516} {\bibfield  {journal} {\bibinfo  {journal} {Phys.
  Rev.}\ }\textbf {\bibinfo {volume} {D94}},\ \bibinfo {pages} {114516}
  (\bibinfo {year} {2016}{\natexlab{e}})},\ \Eprint
  {http://arxiv.org/abs/1608.07585} {arXiv:1608.07585 [hep-lat]} \BibitemShut
  {NoStop}%
\bibitem [{\citenamefont {Lawson}\ \emph {et~al.}(2017)\citenamefont {Lawson},
  \citenamefont {Christ}, \citenamefont {Feng}, \citenamefont {Jüttner},
  \citenamefont {Portelli},\ and\ \citenamefont {Sachrajda}}]{Lawson:2017kxc}%
  \BibitemOpen
  \bibfield  {author} {\bibinfo {author} {\bibfnamefont {A.}~\bibnamefont
  {Lawson}}, \bibinfo {author} {\bibfnamefont {N.~H.}\ \bibnamefont {Christ}},
  \bibinfo {author} {\bibfnamefont {X.}~\bibnamefont {Feng}}, \bibinfo {author}
  {\bibfnamefont {A.}~\bibnamefont {Jüttner}}, \bibinfo {author}
  {\bibfnamefont {A.}~\bibnamefont {Portelli}}, \ and\ \bibinfo {author}
  {\bibfnamefont {C.}~\bibnamefont {Sachrajda}},\ }\bibfield  {booktitle}
  {\emph {\bibinfo {booktitle} {{Proceedings, 34th International Symposium on
  Lattice Field Theory (Lattice 2016): Southampton, UK, July 24-30, 2016}}},\
  }\href {\doibase 10.22323/1.256.0303} {\bibfield  {journal} {\bibinfo
  {journal} {PoS}\ }\textbf {\bibinfo {volume} {LATTICE2016}},\ \bibinfo
  {pages} {303} (\bibinfo {year} {2017})}\BibitemShut {NoStop}%
\bibitem [{\citenamefont {Bai}\ \emph {et~al.}(2017)\citenamefont {Bai},
  \citenamefont {Christ}, \citenamefont {Feng}, \citenamefont {Lawson},
  \citenamefont {Portelli},\ and\ \citenamefont {Sachrajda}}]{Bai:2017fkh}%
  \BibitemOpen
  \bibfield  {author} {\bibinfo {author} {\bibfnamefont {Z.}~\bibnamefont
  {Bai}}, \bibinfo {author} {\bibfnamefont {N.~H.}\ \bibnamefont {Christ}},
  \bibinfo {author} {\bibfnamefont {X.}~\bibnamefont {Feng}}, \bibinfo {author}
  {\bibfnamefont {A.}~\bibnamefont {Lawson}}, \bibinfo {author} {\bibfnamefont
  {A.}~\bibnamefont {Portelli}}, \ and\ \bibinfo {author} {\bibfnamefont
  {C.~T.}\ \bibnamefont {Sachrajda}},\ }\href {\doibase
  10.1103/PhysRevLett.118.252001} {\bibfield  {journal} {\bibinfo  {journal}
  {Phys. Rev. Lett.}\ }\textbf {\bibinfo {volume} {118}},\ \bibinfo {pages}
  {252001} (\bibinfo {year} {2017})},\ \Eprint
  {http://arxiv.org/abs/1701.02858} {arXiv:1701.02858 [hep-lat]} \BibitemShut
  {NoStop}%
\bibitem [{\citenamefont {Bai}\ \emph {et~al.}(2018)\citenamefont {Bai},
  \citenamefont {Christ}, \citenamefont {Feng}, \citenamefont {Lawson},
  \citenamefont {Portelli},\ and\ \citenamefont {Sachrajda}}]{Bai:2018hqu}%
  \BibitemOpen
  \bibfield  {author} {\bibinfo {author} {\bibfnamefont {Z.}~\bibnamefont
  {Bai}}, \bibinfo {author} {\bibfnamefont {N.~H.}\ \bibnamefont {Christ}},
  \bibinfo {author} {\bibfnamefont {X.}~\bibnamefont {Feng}}, \bibinfo {author}
  {\bibfnamefont {A.}~\bibnamefont {Lawson}}, \bibinfo {author} {\bibfnamefont
  {A.}~\bibnamefont {Portelli}}, \ and\ \bibinfo {author} {\bibfnamefont
  {C.~T.}\ \bibnamefont {Sachrajda}},\ }\href {\doibase
  10.1103/PhysRevD.98.074509} {\bibfield  {journal} {\bibinfo  {journal} {Phys.
  Rev.}\ }\textbf {\bibinfo {volume} {D98}},\ \bibinfo {pages} {074509}
  (\bibinfo {year} {2018})},\ \Eprint {http://arxiv.org/abs/1806.11520}
  {arXiv:1806.11520 [hep-lat]} \BibitemShut {NoStop}%
\bibitem [{\citenamefont {Buchalla}\ and\ \citenamefont
  {Buras}(1999)}]{Buchalla:1998ba}%
  \BibitemOpen
  \bibfield  {author} {\bibinfo {author} {\bibfnamefont {G.}~\bibnamefont
  {Buchalla}}\ and\ \bibinfo {author} {\bibfnamefont {A.~J.}\ \bibnamefont
  {Buras}},\ }\href {\doibase 10.1016/S0550-3213(99)00149-2} {\bibfield
  {journal} {\bibinfo  {journal} {Nucl.Phys.}\ }\textbf {\bibinfo {volume}
  {B548}},\ \bibinfo {pages} {309} (\bibinfo {year} {1999})},\ \Eprint
  {http://arxiv.org/abs/hep-ph/9901288} {arXiv:hep-ph/9901288 [hep-ph]}
  \BibitemShut {NoStop}%
\bibitem [{\citenamefont {Misiak}\ and\ \citenamefont
  {Urban}(1999)}]{Misiak:1999yg}%
  \BibitemOpen
  \bibfield  {author} {\bibinfo {author} {\bibfnamefont {M.}~\bibnamefont
  {Misiak}}\ and\ \bibinfo {author} {\bibfnamefont {J.}~\bibnamefont {Urban}},\
  }\href {\doibase 10.1016/S0370-2693(99)00150-1} {\bibfield  {journal}
  {\bibinfo  {journal} {Phys.Lett.}\ }\textbf {\bibinfo {volume} {B451}},\
  \bibinfo {pages} {161} (\bibinfo {year} {1999})},\ \Eprint
  {http://arxiv.org/abs/hep-ph/9901278} {arXiv:hep-ph/9901278 [hep-ph]}
  \BibitemShut {NoStop}%
\bibitem [{\citenamefont {Brod}\ \emph {et~al.}(2011)\citenamefont {Brod},
  \citenamefont {Gorbahn},\ and\ \citenamefont {Stamou}}]{Brod:2010hi}%
  \BibitemOpen
  \bibfield  {author} {\bibinfo {author} {\bibfnamefont {J.}~\bibnamefont
  {Brod}}, \bibinfo {author} {\bibfnamefont {M.}~\bibnamefont {Gorbahn}}, \
  and\ \bibinfo {author} {\bibfnamefont {E.}~\bibnamefont {Stamou}},\ }\href
  {\doibase 10.1103/PhysRevD.83.034030} {\bibfield  {journal} {\bibinfo
  {journal} {Phys.Rev.}\ }\textbf {\bibinfo {volume} {D83}},\ \bibinfo {pages}
  {034030} (\bibinfo {year} {2011})},\ \Eprint {http://arxiv.org/abs/1009.0947}
  {arXiv:1009.0947 [hep-ph]} \BibitemShut {NoStop}%
\bibitem [{\citenamefont {Buras}\ \emph {et~al.}(2005)\citenamefont {Buras},
  \citenamefont {Gorbahn}, \citenamefont {Haisch},\ and\ \citenamefont
  {Nierste}}]{Buras:2005gr}%
  \BibitemOpen
  \bibfield  {author} {\bibinfo {author} {\bibfnamefont {A.}~\bibnamefont
  {Buras}}, \bibinfo {author} {\bibfnamefont {M.}~\bibnamefont {Gorbahn}},
  \bibinfo {author} {\bibfnamefont {U.}~\bibnamefont {Haisch}}, \ and\ \bibinfo
  {author} {\bibfnamefont {U.}~\bibnamefont {Nierste}},\ }\href {\doibase
  10.1103/PhysRevLett.95.261805} {\bibfield  {journal} {\bibinfo  {journal}
  {Phys.Rev.Lett.}\ }\textbf {\bibinfo {volume} {95}},\ \bibinfo {pages}
  {261805} (\bibinfo {year} {2005})},\ \Eprint
  {http://arxiv.org/abs/hep-ph/0508165} {arXiv:hep-ph/0508165 [hep-ph]}
  \BibitemShut {NoStop}%
\bibitem [{\citenamefont {Buras}\ \emph {et~al.}(2006)\citenamefont {Buras},
  \citenamefont {Gorbahn}, \citenamefont {Haisch},\ and\ \citenamefont
  {Nierste}}]{Buras:2006gb}%
  \BibitemOpen
  \bibfield  {author} {\bibinfo {author} {\bibfnamefont {A.~J.}\ \bibnamefont
  {Buras}}, \bibinfo {author} {\bibfnamefont {M.}~\bibnamefont {Gorbahn}},
  \bibinfo {author} {\bibfnamefont {U.}~\bibnamefont {Haisch}}, \ and\ \bibinfo
  {author} {\bibfnamefont {U.}~\bibnamefont {Nierste}},\ }\href {\doibase
  10.1007/JHEP11(2012)167, 10.1088/1126-6708/2006/11/002} {\bibfield  {journal}
  {\bibinfo  {journal} {JHEP}\ }\textbf {\bibinfo {volume} {0611}},\ \bibinfo
  {pages} {002} (\bibinfo {year} {2006})},\ \Eprint
  {http://arxiv.org/abs/hep-ph/0603079} {arXiv:hep-ph/0603079 [hep-ph]}
  \BibitemShut {NoStop}%
\bibitem [{\citenamefont {Brod}\ and\ \citenamefont
  {Gorbahn}(2008)}]{Brod:2008ss}%
  \BibitemOpen
  \bibfield  {author} {\bibinfo {author} {\bibfnamefont {J.}~\bibnamefont
  {Brod}}\ and\ \bibinfo {author} {\bibfnamefont {M.}~\bibnamefont {Gorbahn}},\
  }\href {\doibase 10.1103/PhysRevD.78.034006} {\bibfield  {journal} {\bibinfo
  {journal} {Phys.Rev.}\ }\textbf {\bibinfo {volume} {D78}},\ \bibinfo {pages}
  {034006} (\bibinfo {year} {2008})},\ \Eprint {http://arxiv.org/abs/0805.4119}
  {arXiv:0805.4119 [hep-ph]} \BibitemShut {NoStop}%
\bibitem [{\citenamefont {Arthur}\ \emph {et~al.}(2013)\citenamefont {Arthur}
  \emph {et~al.}}]{Arthur:2012yc}%
  \BibitemOpen
  \bibfield  {author} {\bibinfo {author} {\bibfnamefont {R.}~\bibnamefont
  {Arthur}} \emph {et~al.} (\bibinfo {collaboration} {RBC, UKQCD}),\ }\href
  {\doibase 10.1103/PhysRevD.87.094514} {\bibfield  {journal} {\bibinfo
  {journal} {Phys. Rev.}\ }\textbf {\bibinfo {volume} {D87}},\ \bibinfo {pages}
  {094514} (\bibinfo {year} {2013})},\ \Eprint {http://arxiv.org/abs/1208.4412}
  {arXiv:1208.4412 [hep-lat]} \BibitemShut {NoStop}%
\bibitem [{\citenamefont {Feng}\ \emph {et~al.}(2010)\citenamefont {Feng},
  \citenamefont {Jansen},\ and\ \citenamefont {Renner}}]{Feng:2009ij}%
  \BibitemOpen
  \bibfield  {author} {\bibinfo {author} {\bibfnamefont {X.}~\bibnamefont
  {Feng}}, \bibinfo {author} {\bibfnamefont {K.}~\bibnamefont {Jansen}}, \ and\
  \bibinfo {author} {\bibfnamefont {D.~B.}\ \bibnamefont {Renner}},\ }\href
  {\doibase 10.1016/j.physletb.2010.01.018} {\bibfield  {journal} {\bibinfo
  {journal} {Phys. Lett.}\ }\textbf {\bibinfo {volume} {B684}},\ \bibinfo
  {pages} {268} (\bibinfo {year} {2010})},\ \Eprint
  {http://arxiv.org/abs/0909.3255} {arXiv:0909.3255 [hep-lat]} \BibitemShut
  {NoStop}%
\bibitem [{\citenamefont {Christ}\ \emph
  {et~al.}(2015{\natexlab{b}})\citenamefont {Christ}, \citenamefont {Feng},
  \citenamefont {Martinelli},\ and\ \citenamefont
  {Sachrajda}}]{Christ:2015pwa}%
  \BibitemOpen
  \bibfield  {author} {\bibinfo {author} {\bibfnamefont {N.~H.}\ \bibnamefont
  {Christ}}, \bibinfo {author} {\bibfnamefont {X.}~\bibnamefont {Feng}},
  \bibinfo {author} {\bibfnamefont {G.}~\bibnamefont {Martinelli}}, \ and\
  \bibinfo {author} {\bibfnamefont {C.~T.}\ \bibnamefont {Sachrajda}},\ }\href
  {\doibase 10.1103/PhysRevD.91.114510} {\bibfield  {journal} {\bibinfo
  {journal} {Phys. Rev.}\ }\textbf {\bibinfo {volume} {D91}},\ \bibinfo {pages}
  {114510} (\bibinfo {year} {2015}{\natexlab{b}})},\ \Eprint
  {http://arxiv.org/abs/1504.01170} {arXiv:1504.01170 [hep-lat]} \BibitemShut
  {NoStop}%
\bibitem [{\citenamefont {Blum}\ \emph {et~al.}(2016)\citenamefont {Blum} \emph
  {et~al.}}]{Blum:2014tka}%
  \BibitemOpen
  \bibfield  {author} {\bibinfo {author} {\bibfnamefont {T.}~\bibnamefont
  {Blum}} \emph {et~al.} (\bibinfo {collaboration} {RBC, UKQCD}),\ }\href
  {\doibase 10.1103/PhysRevD.93.074505} {\bibfield  {journal} {\bibinfo
  {journal} {Phys. Rev.}\ }\textbf {\bibinfo {volume} {D93}},\ \bibinfo {pages}
  {074505} (\bibinfo {year} {2016})},\ \Eprint {http://arxiv.org/abs/1411.7017}
  {arXiv:1411.7017 [hep-lat]} \BibitemShut {NoStop}%
\end{thebibliography}%
\end{document}